# Complete Separation of the 3 Tiers

*Divide and Conquer*

### Walter Pidkameny

AT&T Chief Security Office
180 Park Ave D179
Florham Park, NJ 07932, USA
1 (973) 360-5828

wp@att.com

### Victor Sadikov

AT&T Information Technology Unit
200 S Laurel Ave A5-2D20
Middletown, NJ 07748, USA
1 (732) 420-7453

vic@att.com

## ABSTRACT

Most Java applications, including web based ones, follow the 3-tier architecture. Although Java provides standard tools for tier-to-tier interfaces, the separation of the tiers is usually not perfect. E.g. the database interface, JDBC, assumes that SQL statements are issued from the application server. Similarly, in web based Java applications, HTML code is assumed to be produced by servlets. In terms of syntax, this turns Java source code into mixtures of languages: Java and SQL, Java and HTML. These language mixtures are difficult to read, modify, and maintain.

In this paper we examine criteria and methods to achieve a good separation of the 3 tiers and propose a technique to provide a clean separation. Our proposed technique requires an explicit Interface and Data Definitions. These allow isolation of the back-end, application server, and front-end development. The Definitions also enable application design in terms of aggregated data structures. As a result significant amounts of auxiliary code can be generated from the Definitions, enabling the developers to concentrate on the business logic. By and large the proposed approach greatly facilitates development and maintenance, and overall improves the quality of the products.

## Categories and Subject Descriptors

D.2.11 [**Software Engineering**]: Software Architectures – *languages, patterns.* D.2.12 [**Software Engineering**]: Interoperability – *interface definition languages.* D.1.5 [**Programming Techniques**]: Object-oriented Programming. D.2.13 [**Software Engineering**]: Reusable Software – *reusable libraries, reuse models.* D.3.3 [**Programming Languages**]: Language Constructs and Features – *classes and objects, control structures, data types and structures, frameworks, inheritance.* H.5.4 [**Information Interfaces and Presentation**]: Hypertext/Hypermedia – *architectures.* I.7.2 [**Document and Text Processing**]: Document Preparation – *markup languages.*

## 1. INTRODUCTION

There are many technologies for developing a web based application. The fact that there are so many implies there are some common problems, which are identified and addressed by a number of developers. All the web technologies have drawbacks, and none of the technologies is clearly superior to all the others.

Usually a technology is based on an original solution which seems so beneficial that it becomes an industry standard before the technology is refined or even logically completed. We do not claim that our approach is absolutely innovative. Rather, we examined a number of popular web technologies and identified the problems they were intended to address. Based on our analysis we selected the best solution, integrated it in our system, and consistently followed it in all similar cases. Our survey covers twelve technologies.

### 1.1 Servlets

Java Servlet technology provides Web developers with a platform-independent, consistent mechanism for building interactive Web applications. Usually, a servlet accepts CGI parameters and generates an HTML page which defines the look-and-feel of the web application. Servlets are implemented in **Java**. Java is the most convenient language to implement the business logic because of its many benefits, including access to the entire family of Java APIs, portability, performance, reusability, and crash protection.

Changing the look-and-feel of a current application is a rather frequent task, but one which should not require any development on the application tier. Therefore, it is very desirable to separate the implementation of business logic from the presentation. Unfortunately, regular servlets do not create this separation, because it is assumed that HTML code needs to be literally included in the servlet output statements. HTML code is very difficult to read by itself, even more so when HTML code is interspersed within a Java servlet.

### 1.2 JSP

JSP technology is intended to compensate for some drawbacks of HTML. E.g. being a data markup language, HTML lacks all programming constructions, and does not support any reusable code, loops, conditional branches, or global variables. JSP re-introduced **programming constructions** and dynamic data into HTML code. The cost for this benefit is a JSP page that is an ugly mixture of two different languages, HTML and Java.

In terms of the application tier development, JSP allow adding business functionality to a web page. However, since the presentation is in the same file as the business logic, the developer needs to be competent in both HTML and Java to be able to maintain or change the JSP application.

## 1.3  Template Engines

This approach assumes that the HTML code should be localized in external files. The application produces dynamic data in a form of a hash table, and a **template** engine uses the dynamic data to populate the resulting web page.  It is essential that the template engine supports basic text processing operations, loops, branching, and enclosed template calls.

This approach provides good separation between the presentation and the business logic.  It also facilitates web site development by improving HTML with programming constructions and dynamic data support. The efficiency and convenience of this approach depends on the integration of HTML and the particular template language.  Regular macro generators and preprocessors are not good at dealing with HTML. Another option is **XSL**, which in spite of its clumsy syntax is highly compatible with HTML.

## 1.4  JDBC

Java Database Connectivity API (JDBC) is an industry **standard**, intended to provide *database-independent* connectivity between the Java programming language and a *wide range* of databases. Actually, JDBC only claims to work with SQL-based databases. Since SQL databases are most common, even with such a limitation, the actual range of supported databases is still wide.

The question of database independence is more subtle.  JDBC is database independent in the sense that it can access whatever database platform (MySQL, Oracle, etc.) being used. Unfortunately, due to the different database implementations and various SQL dialects, specific JDBC/SQL code may cause extra *database dependence*.  However, database incompatibility is not usually a major issue since switching from one database to another is both a rather rare task and one that requires a fair amount of programming effort any way.

A more important and more frequent task is switching from one database schema to another within the same database platform. Good database independence assumes that this task should involve absolutely no development on the application tier.

Unfortunately, this is not the case with JDBC.  Java code using JDBC literally contains SQL statements which *are* database schema dependent, because they specify which fields in which tables should be retrieved, updated or deleted.  They also explicitly depend on the table relationships.  Splitting tables, joining tables, changing table references or indices, significantly affect the SQL statement pattern.

## 1.5  SQL Strings (Q Files)

This approach assumes that **SQL statements** should be placed in external files separate from the JDBC statements.  This approach provides slightly better separation between the application server and the back end.  Unfortunately, even though the Q files can be maintained by a database developer, a Java developer is still involved in programming JDBC statements, constructing SQL statements, receiving and parsing results.

## 1.6  Stored Procedures

Stored procedures isolate enclosed SQL code and provide a fixed interface defined by the **stored procedure** signatures.  The underlying implementation can change freely with no impact to the rest of the application as long as the stored procedure signature and functionality are the same.  Thus it is very beneficial to wrap even a single SQL statement into a stored procedure, because it makes the call independent of a particular database schema.  Fortunately, stored procedures can be called from JDBC statements and return not only scalar values but also result sets.

## 1.7  Postgres

The Postgres database project aims to add complete support for **types**. Its features include the ability to define types, as well as the ability to fully describe relationships.  The relationships are defined as a number of rules belonging to the database.  The required data are retrieved from related tables and aggregated to application objects at the database level.

## 1.8  Connection Management

Database connection management requires special attention.  The JDBC based application needs to properly open and even more importantly, close database connections.  Furthermore, connections and result set cursors should be closed in all possible cases, specifically when SQL exceptions occur.  So, it is recommended to use one class to mange all connections in the same pattern.  Particular classes dealing with the database can use methods of the connection management class.  Following OOP approach, connection management may be implemented in a **base database class**, with particular database dealing classes being extended from it.

## 1.9  EJBs

Enterprise Java **Beans** were intended to handle common back-end concerns, such as persistence, transactional integrity, and security in a standard way. Unfortunately, EJBs are not easy to develop and use, and there is no way to use just a subset of EJB functionalities.  Any usage of EJBs assumes following a complicated programming model.   The EJB server handles distributed object, distributed transactions management, and other services for enterprise beans. The EJB containers interface with enterprise beans at runtime, implementing the session bean contract or the entity bean contract. The application enterprise beans require three classes for every session bean and four classes for an entity bean.  In many cases an EJB solution may be more complex and require more development than otherwise necessary.

## 1.10  CORBA

Common Object Request Architecture and infrastructure (CORBA) is used to make a computer application work together over networks with other applications on different computers, running under different operating systems, and written in different programming languages.  This ambitious task is accomplished by *separation of interface from implementation* and *Stub/ORB/Skeleton architecture*.

The **interface** to each object is defined very strictly and independently of any programming language.  Clients access objects only through their advertised interfaces, invoking only those operations that the object exposes through its IDL interface. The *implementation* of an object - its running code, and its data - is hidden from the rest of the system. Stub/ORB/Skeleton Architecture assumes that Stubs and Skeletons serve as proxies for clients and servers, respectively.

A considerable advantage of CORBA is that client stubs and object skeletons **are compiled** from IDL. IDL compilation facilitates object implementation and usage, and guarantees that client stubs and object skeletons will have no trouble meshing perfectly, even if the two are compiled into different programming languages.

## 1.11 WSDL

WSDL is a specification defining an interface between the HTTP service requestor and the service provider, in terms of messages, services, ports, network endpoints, etc. WSDL takes into account a large number of software vendors. The specification is platform and language-independent and is written in a common **XML** grammar.

## 1.12 Struts

Struts refers to the three-tier architecture as a Model-View-Controller framework. When applied to standard Java EE web applications, the *model* is the application logic that interacts with a database; the *controller* is responsible for passing information between view and model; and the *view* comprises HTML pages presented to the client.

The goal of Struts is to cleanly **separate** the *model* from the *view* and the *controller*. To this end Struts provides a framework consisting of the following components. Actions are the requests from the client sent to the controller. The Controller receives requests and calls corresponding Action Classes. Action Classes interact with the application specific model code. ActionForward is a string returned from the model, which tells the controller what output page should be returned to the client. Special Java Beans are used to pass information between model and view. A custom JSP tag library allows reading and writing the content of these beans from the presentation layer without the need for any embedded Java code. The Struts Framework consists of several hundred Java classes and covers all three tiers.

Unfortunately, Struts does not substitute the technologies described above. On the contrary, the standard is designed to support all features in all possible technologies, with no regard to coherency of the implementation. As the result, Struts has to deal with, and compensate for all the drawbacks. This makes Struts overblown and overly complex. A lightweight framework can be a better solution to most if not all problems.

## 1.13 Spring MVC

Spring MVC is recently gaining its popularity due to ability to address many technical challenges of web development. Spring Framework consistently follows the MVC concept and is a good light-weight alternative to Struts.

Spring Framework provides high independence of the request handling layer, presentation and the model. According to Spring MVC, the business logic is localized in the Controller class. The Spring MVC engine is responsible for mapping a client request to the appropriate Controller, and for resolving and rendering the View back to the client based on a set of given workflow rules called the application context.

The implementation of Spring MVC makes extensive use of runtime *interpretation* and Java reflection. Interpretation provides excessive flexibility of source code but also results in ignoring important compatibility information otherwise available at the compile time. Thus, the consistency between hand written model classes and the application context is not guaranteed automatically; and should be manually maintained by the application developers. Another weakness of Spring MVC is a lack of a high-level abstraction for web-based applications, which results in tight coupling of all interfaces to the Servlet API.

## 2. DESIGN

Having analyzed common problems and popular solutions we offer the following 3-tier architecture. The Front End is a web based client or a stand alone Java client. The Business Logic is implemented on a web server, for instance as a number of servlets or C++ CGI executables. The Back End is an SQL database, such as MySQL or Oracle. We will use XML over HTTP as a connection between the application server and the front end; and JDBC as a connection between the application server and the database. These are our preferences; so far – nothing new.

But in addition we will completely separate the three tiers by defining the front and back interfaces in a concise and language/layer independent form of XML clauses. These API definitions become the complete and final agreements between the GUI, server, and database. Ideally, such APIs will mean the same things for the developers of different layers, but in different terms. Thus, a web client developer is interested in what CGI parameters the web server expects and what XML data the server will issue back; but he is not interested in the internal representation of these parameters or data on the server side. Likewise, the database developer is interested in which stored procedures with which signatures he needs to implement, but he is hardly interested in which wrapper classes will be used to access his stored procedures and manage connections. Finally the application server developer will find convenient to use beans and wrapper classes with setter and getter methods for processing aggregated data, issuing screens and accessing stored procedures.

We claim our tool can support the interfaces as described above. The implementation is based on the idea that the API definitions can be used to generate servlet stubs and skeletons, screen classes, stored procedure wrapper classes, and aggregated data beans used throughout the whole application. The generated auxiliary classes will guarantee that the front end, application server, and the back end are consistent and can work together.

## 3. THE DETAILS

We start with introducing the technique of API definitions, and then we will consider all parts of a complete example application. The API definitions describe the application data and the tier to tier interfaces. The simplest types of application data are linear beans containing fields of predefined terminal types. More complicated data structures are composite beans consisting of both predefined types and other beans. Composite beans can also include vectors of beans or predefined types. Client and database interfaces define what calls can be made from a tier to tier and what data structures should be passed to, and be received from these calls. E.g. a general Java interface allows making a call from a client to server. The client instantiates the server wrapper class, sets the input data, makes a call to the process method, and gets the output data. The server ought to implement the process method. A special case of a tier to tier interface is a call of a stored procedure wrapper class. The database interface includes the definitions of the stored procedure signatures. The application server makes the call in exactly the way described above. But the stored procedure wrapper class contains no manually written implementation. The whole class is generated. The generated process method opens a database connection, makes a JDBC call, retrieves the results, and closes the connection. All the back end logic is shifted to the stored procedure and is implemented entirely in SQL.

### 3.1 APIs

We begin design and development of an application by examining application data. Assuming our example application deals with customers, we will probably need such data as first name, last name, phone number, street, city, zip and state. As we discussed earlier, we prefer to think in terms of aggregated data structures. In our example, they are Personal Information and Address. Personal Information consists of the First and Last Names. The Address contains City, Zip, State, etc. The definitions for these beans will look like the following.

```
<bean name="Person">
        <param name="FirstName" type="CcName">
        <param name="LastName" type="CcName">
</bean>

<bean name="Address">
        <param name="Street"  type="CcString">
        <param name="City"    type="CcName">
        <param name="State"   type="CcString">
        <param name="Zip"     type="CcNumber">
</bean>
```

Here CcString, CcName, and CcNumber are predefined types. They will be useful for validating and formatting the inputted data.

### 3.2 Beans

The definitions of the Person and Address beans will result in creating the Java classes "example.common.bean.Person" and "example.common.bean.Address" with the following setter and getter methods:

```
package example.common.bean;

class Person extends CcBean
{
        public CcName  FirstName = new CcName();
        public CcName  LastName = new CcName();
        public void setFirstName(String x) ...
        public String getFirstName() ...
        public void setLastName(String x) ...
        public String getLastName() ...

        public String validate() ...
}
class Person extends Address
{
        public void setStreet(String x) ...
        public String getStreet() ...
        public void setCity(String x) ...
        public String getCity() ...
        public void setState(String x) ...
        public String getState() ...
        public void setZip(String x) ...
        public String getZip() ...
}
```

The beans are rather powerful. They have methods to validate, format, and encrypt each field and the whole bean. They also have convenient methods to convert themselves to and from Java hash tables, JDom elements, and JDBC result sets. They just require predefined types and the base class to be fully functional. In addition, they can be used and be useful even outside our tool.

### 3.3 Common Types

The beans rely on the predefined types, which are responsible for validating and formatting data. The tool provides a set of predefined types for common data like strings, names, numbers, etc. The types form a hierarchy, starting from the CcString (5.3.3) type, which has almost no restrictions. The CcName type extends from CcString. CcName should contain only letters. All letters should be converted to the lower case. When displayed, the first letter should be capitalized. Another predefined type is CcNumber. CcNumber consists of only digits. There are also types for particular numbers. CcPhone, which extends from CcNumber and meets all its limitations, should contain exactly 10 digits. When entered, CcPhone may contain brackets, spaces, and dashes. When displayed, CcPhone is formated to a string like "(123)456-7890." Another CcNumber is CcZip. CcZip also extends from CcNumber. CcZip should contain 5 or 9 digits. There are also types for credit card and SSN numbers. The predefined set of types can be easily extended, if needed.

### 3.4 Composite Beans

In the previous sections we discussed that it is convenient and natural to define beans and then to design, develop, and modify the application at hand in terms of beans rather than in terms of particular fields. As a result the beans become just new data types. So it will be reasonable to allow including one bean into another to construct more aggregated structures. In our example we can define Customer as a bean consisting of Personal Information and Address.

```
<bean name="Customer">
        <bean name="Person" type="Person">
        <bean name="Address" type="Address">
</bean>
```

It would be possible to hide in the API definition the fact that fields of this bean are beans themselves. But dealing with the fields of a bean the developer must always explicitly know whether he is dealing with a terminal type or an included bean, so we introduce the statement tag "bean."

```
class Customer extends CcBean
{
        public void setPerson(Person x) ...
        public Person getPerson() ...
        public void setAddress(Address x) ...
        public Address getAddress() ...
}
```

### 3.5 Delivered Beans

The Object Oriented approach would not be perfectly followed if we restricted our beans to extend one from another. In fact, we did come across such a situation in our programming practice. We had to send messages to our customers. A message was a bean. But there also were different types of messages: email messages, recurring messages, broadcast messages, etc. They had a slightly different set of fields, but they all extend from the regular message. The following are our APIs.

```
<bean name="EmailMessage" extends="Message">
        <param name="EmailAddress" type="CcEmail">
</bean>
```

This definition produces the class "EmailMessage."

```
class EmailMessage extends Message
{
        public void setEmailAddress(String x) ...
        public String getEmailAddress() ...
}
```

The EmailMessage class extends from the Message class, so it includes all fields and methods of the base class, plus its own setter and getter methods. General methods (like validate, format, and convert) work with the base class first, and then with the field of the extended bean.

## 3.6  Vectors

It is not typical for a bean to contain a vector of other beans. But sometimes applications need to deal with vectors of beans. To this end we allow a vector of beans in all places where a scalar bean or type is allowed. The corresponding API syntax is the following.

```
<bean name="EventList">
        <vector name="Event"  type="EventItem">
</bean>
```

The result EventList bean contains a vector (5.3.6) of the EventItem beans.

```
class EventList extends CcBean
{
        public vector EventVector = new Vector();

        public void setVectorOfEvent(Vector x) ...
        public Vector getVectorOfEvent() ...
        public void addEvent(EventItem x) ...
}
```

This example EventList bean can be used as a wrapper for a vector of EventItems.

The vector can be set and get as a whole. In addition, there is a method to add an element of the defined type to the vector. All elements of the vector participate in the general bean operations like validate, convert, load, etc. One bean can contain more than one vector. The fields of vectors are common when defining interfaces because it is quite usual to get a list of records from the database and pass it to the client to display.

## 3.7  Java Interfaces

What was shown above reflects our understanding of a pure Java interface. On the client side the interface should look like a server proxy class with a number of setter and getter methods, and the functional method, something like "process," or "execute." The client instantiates the class, sets input parameters, calls the process method, and retrieves output parameters by means of the getter methods. The client is not interested in the implementation of the server class, just in the interface definition.

```
ServerX server = new ServerX();
server.setInputA(a);
server.setInputB(b);
server.process();
OutputC c = server.getOuputC();
OutputD d = server.getOuputD();
```

This example interface can be defined as the following.

```
<transaction name="ServerX">
        <request>
                <bean name="InputA"  type="InputA">
                <bean name="InputB"  type="InputB">
        </request>
        <response>
                <bean name="OutputC"  type="OutputC">
                <bean name="OutputD"  type="OutputD">
        </response>
</transaction>
```

Such an interface can be used between a Java server and a stand alone Java GUI. We have already shown how to call it from the client.

## 3.8  Business Logic

Implementation on the server side is a bit tricky. The tool can generate an example skeleton, but then it is the responsibility of the developer to add appropriate business logic.

```
class ServerX_Skeleton extends CcServlet
{
        public void setInputA(InputA x) {a=x;}
        public void setInputB(InputB x) {b=x;}
        public OutputC getOutputC() {return c;}
        public OutputD getOutputD() {return d;}

        public void process()
        {
                process(a, b, c, d);
        }

        public void process(InputA a, InputB b, OutputC c, OutputD d)
        {
                error("business logic should be implemented here");
        }
}
```

The trick is to use OOP to avoid changes in the generated file. The business logic could be implemented in a separate class, which extends from the generated skeleton. The developer needs to overwrite the process method with the one containing the actual business logic.

```
class ServerX extends ServerX_Skeleton
{
        public void process(InputA a, InputB b, OutputC c, OutputD d)
        {
                // actual business logic goes here
        }
}
```

## 3.9  Server Call

Let's track a server call. The client makes a call to the server proxy. The proxy packs parameters into an XML structure and makes an HTTP request to the web server. The web server invokes the appropriate servlet, which is a skeleton part of the interface. The servlet parses XML parameters and calls its own "process" method, which is overwritten by the developer with a method implementing the required business logic. Then the control returns to the skeleton servlet. The servlet packs output parameters into an XML structure and returns it as the HTTP response. The server proxy class parses the returned parameters from the response and returns them to the client.

Notice, that all the components involved are generated from the API definition. And at every point they know what parameters (what names and what types) to pass and to expect.

## 3.10  Exception Handling

The above scenario assumes that everything goes well. But if something goes wrong, and the server fails to return a meaningful result, the server should inform the client of an error. Usually there are a great number of different errors, but the client may be interested in distinguishing just a few. Besides, the error may not originate on the client side, it might be caused by an error on an external system, etc. To deal with errors, Java language provides the technique of **exceptions**. An exception can keep a chain of root causes. Being classes, exceptions can form a hierarchy reflecting different errors. The client can catch a particular exception, distinguish just some nodes of an exception hierarchy, or process all exceptions in the same way.

We consider exceptions a convenient way to handle errors. The Summer Framework supports exceptions. If an exception is thrown on the server side, the skeleton servlet will catch it and convert the whole exception chain to a special XML structure. The server proxy class understands the exception XML and recreates the final exception as well as the exception chain. This is implemented by means of Java reflection. Then it throws the final exception including the exception chain as the cause.

The tool also provides some basic exception hierarchy. Everything starts with CcException and then goes to CcApplicationError or CcSystemError. CcSystemError can be further detailed as CcCommunicationError, or CcUnavailable, or CcChainError, etc.

## 3.11  Stored Procedure Wrappers

We recommend using stored procedures in the database interface. An API definition for a stored procedure is similar to the API definition for a Java interface. Even with limitations (5.3.11.) the interface is still rather powerful. E.g. a stored procedure can return a vector of application beans as a cursor to a result set. Consider the following example definition.

```
<procedure name="GetPeople">
    <request>
        <param name="Name" type="CcString">
    </request>
    <response>
        <vector name="Result" type="Person">
    </response>
</procedure>
```

A call to a stored procedure is made from the middle tier, where the wrapper for the particular stored procedure is instantiated. Then the necessary input parameters are set by the corresponding setter methods, and the call is executed. The execution of the call involves JDBC communication between the stored procedure wrapper and the database server. The wrapper class takes care of the getting a spare connection from the pool, preparing a callable statement, reading returned parameters, closing JDBC statements and cursors, and returning the connection to the pool. Then the output parameters can be accessed by the getter methods. Some input parameters can be vectors of strings, some output parameters can be vectors of beans, with everything being exactly as defined in the API.

The invocation of a stored procedure code is similar to the invocation of the Java server call, but the database implementation is quite different from the server implementation. It is even written in SQL language, not in Java. Notice too that nowhere (but the code generated from APIs) are these two languages mixed.

Once again, the tool will also generate a stub for database implementation, which will be just a stored procedure signature.

```
procedure GetPeople_Proc
(
        Name  in varchar2,
        Result  out cursortype
);
```

The body of the stored procedure contains all database dependent code, and should, of course, be manually written.

```
Begin
        open Result for
        SELECT
                table_p.first  as  FirstName,
                table_p.last  as  LastName
        FROM table_p
        WHERE  table_p.last  LIKE  Name;
end GetPeople_Proc;
```

Returning a result set with named parameters "FirstName" and "LastName" will allow the stored procedure wrapper to construct a vector of the Person beans. This information can be found in the bean API definitions.

## 3.12  Web Client Interface

Generally, one should keep the interface for a web based client exactly the same as the one for a stand alone Java client. The only difference is that now the application server decides which screen to display. The responsibility of the client will be to render a screen template and to populate it with the dynamic data provided by the application server. This results in decoupling the request and the response parts of the API definitions.

## 3.13  Request Handlers

The request part of the interface will look like the following.

```
<request name="HandlerX">
        <bean name="InputA"  type="InputA">
        <bean name="InputB"  type="InputB">
        <bean name="Customer"  type="Customer">
</request>
```

This example means that the servlet managing this request accepts three beans as parameters. A request handler servlet always returns a screen, defined in the API definitions. The skeleton for this servlet will look like the following.

```
class HandlerX_Base extends CcRequestHandler
{
        public CcScreen process(InputA a, InputB b, Customer c)
        {
                error("business logic should be implemented here");
                return null;
        }
}
```

And once again, instead of being put directly here, the business logic should be implemented in the extended class.

```
class HandlerX extends HandlerX_Base
{
        public CcScreen process(InputA a, InputB b, Customer c)
        {
                // actual business logic goes here
                return new CcErrorScreen("to be implemented");
        }
}
```

## 3.14   Request Parameters

We added the Customer bean to our example to demonstrate how to pass bean parameters to a servlet. The bean is named Customer. Looking at its definition, we find that it contains two sub-beans named Person and Address. The sub-beans include plain fields named FirstName, LastName, Street, etc.

According to our naming convention for a web-based GUI, these parameters should be referred to as "Customer.Person.FirstName," "Customer.Person.LastName," "Customer.Address.Street," etc. I.e. the path names of included beans are separated by dots. Fortunately CGI supports names with dots.

This convention allows using several instances of the same beans. E.g. if we need to pass a vendor information to the same servlet, we can use names like "Vendor.Address.Street," as long as they are correctly defined in the APIs.

The request handler base class will take care of parsing CGI parameters, instantiating and initializing the expected beans, and passing them to the overwritten process method. It will also take care of the HTTP cookies (although we do not go into further detail here.)

## 3.15   Screens

Screens represent the dynamic data issued by request handler servlets. They are defined as a decoupled response part of the API definitions.

```
<screen name="ScreenX">
    <bean name="InputA" type="OutputC">
    <bean name="InputB" type="OutputD">
    <bean name="Customer" type="Customer">
</screen>
```

Screens are very similar to beans, but we do not allow a request handler to issue an arbitrary bean, because the web client probably does not expect it.

Having processed the input parameters, the request handler process method instantiates, populates and emits an appropriate screen. How to process the input data and what screen to emit is controlled by the business logic. Then the request handler base class converts the screen bean and all its sub-beans and vectors into XML format and returns it as an HTTP response. The XML could look like the following.

```
<Screen name="ScreenX">
    <Customer>
        <Person>
            <FirstName>Thomas</FirstName>
        </Person>
    </Customer>
</Screen>
```

## 3.16   Rendering a Screen

In general terms, a template engine will be responsible for finding the template corresponding to the emitted screen, rendering the template, and populating it with the dynamic data. This is a clean solution, i.e. there is no HTML or other language code inserted into the application server Java code and there is no Java code on the client side. The client server interface strictly defines what CGI parameters and what XML structure should be passed between the tiers. Dynamic data are completely isolated from the look and feel information. This accomplishes our goals, but the success of our tool will probably depend on how successfully we can facilitate web client development.

For more information on a successful implementation of a web based client intended to work with our tool please see "OOML: Structured Approach to Web Development" by Francisco and Sadikov (2008.) Below we will just outline some key principles.

## 3.17   XSL Templates

We use XSL transformation for rendering HTML pages and populating them with dynamic data. XSL transformation allows creating web pages in terms of web page structures. E.g. a web page usually consists of a header at the top, a footer at the bottom, a navigation column to the left, a promotion column to the right and content in the center. The content can consist of a title and a form. The form may consist of input fields with names. This structure is quite obvious when looking at the web page, but it is completely hidden in the HTML code. When using XSL transformation the web page structure can be extracted into a separate file in concise XML format.

```
<Screen name="Search">
    <Header>Example Application</Header>
    <Navigation>
        <MenuItem href="home.html">Home</MenuItem>
        <MenuItem href="search.html">Search</MenuItem>
    </Navigation>
    <Content>
        <Title>Customer Search</Title>
        <Form href="results.html">
            <InputField prop="Customer.Address.City">City:</InputField>
        </Form>
    </Content>
    <Promo/>
    <Footer>2008</Footer>
</Screen>
```

The web page structure describes only significant functional elements of the target web page. All presentation details are shifted to XSL templates. Web page structures form an extra layer of web development. All functional changes, such as modifying a menu item or adding a field to the form, can easily be done at this level. On the other hand, all changes to look-and-feel are done in the XSL templates which will affect all pages of the site.

The tool includes an XSL template library supporting a complete set of functional elements and templates to work with web page structures and dynamic data. The root template is responsible for extracting the name of the requested screen from the dynamic data, finding a corresponding web page structure, and applying to it specified templates. Layout templates allow placing web elements in specific places. Decoration templates format static text and titles. Data templates access, select, and display dynamic data. Input elements initialize input fields with the default values taken from dynamic data, and put entered values into conventionally named CGI parameters.

The library supports branches and loops depending on static and dynamic data. It also allows creating and using additional templates as macros.

All decoration elements are controlled by global values enabling easy adjustment of the site to a particular color scheme, font faces and sizes. If more serious look and feel changes are required, the XSL templates can be modified without affecting the site functionality.

## 4. THE SUMMER MVC

The Summer MVC tool is provided as a main Ant script, a code generator implemented as a number of XSL files, and a library of auxiliary run time Java classes. To use the tool, the application data structures and tier interfaces should be defined in application API files. Then the generator is invoked to generate application common beans, screens and database wrapper classes. Next, the generated classes are compiled and packed into a Java archive. After that the beans and interface classes become available for the application server Java project.

All the steps, including compilation of Java project and deployment of project classes along with a library of the generated common beans, interfaces, and auxiliary Java classes, can be done by executing just one Ant target. The tool can be easily integrated into Java development IDE, such as JDeveloper or Eclipse.

### 4.1 A Simple Example

Regrettably, the limited size of this paper does not allow producing a detailed example. Fortunately, this is not necessary either, because we have already discussed all the components in the previous sections. Here we will just outline the recommended development of an application. It starts with defining aggregated application data structures, such as Customer, Personal Information and Address (see Section 3.1.) Then the interface definition should be created to enumerate and define all possible requests from the GUI to the application server and from the application server to the database stored procedures (see Sections 3.7, 3.13, 3.15, and 3.11.) A simple application can process just one request, e.g. to find a customer. The GUI passes the desired name or phone number to the application server as described in section 3.14. The trivial request handler involves no specific business logic, but just forwards the input parameters to the corresponding stored procedure and sends the result back. Interface definitions for this case include one request handler, and one stored procedure signature.

From this point, the GUI, application server, and database developers can work independently. A standalone GUI can be written in Java and will be able to call the defined request handlers on the server side as described in Section 3.9 and 3.10. A web-based GUI can be entirely written in a markup language as described in "OOML: Structured Approach to Web Development" by Francisco and Sadikov (2008.) In both cases the application server is realized as a set of generated servlets corresponding to the defined request handlers. The business logic is implemented in Java and will reside in the middle tier, see Section 3.8. The application server will use a set of convenient wrapper classes to access the stored procedures. The stored procedures involved will be implemented in SQL, as described in Section 3.11.

### 4.2 Projects

We have successfully applied the Summer MVC to a number of existing projects mostly involved in self and agent based customer care web applications. The tool allowed us to make the projects more understandable, and the code more readable. This significantly reduced development and maintenance time and cost.

### 4.3 Evolution

As stated above the Summer Framework does not require a new project or a one time switch from an old technology. On the contrary, the tool can be gradually applied to an existing project.

Among the few projects that have utilized the tool, only one was implemented with completely new code. All the others were existing projects and we had no luxury to switch a whole project at once. So we had to apply the Summer MVC to just those parts of the projects where new features or modifications were required. When following this approach, it took months to completely switch the projects to the introduced technology. As each feature was ported, the applications became more comprehensible, maintainable, and reliable.

The Summer Framework itself also underwent some evolution, but all the changes were to fix minor bugs, and to introduce a couple of new, more convenient functions. And the best thing was that fixing a bug just in one place, removed that bug from all relevant functions in all our projects implemented with the tool.

## 5. REMARKS

This section intentionally contains sub-sections 5.3.3, 5.3.6, and 5.3.11 only.

### 5.3.3 Java String

As we described in section 3.3, the common type class hierarchy starts from the CcString class. This appears to be an omission in the design. It is a normal expectation for the class hierarchy to start with the most appropriate standard Java class. Further, the name of the base root class suggests that its functionality is quite close to the predefined type/class String. A proprietary String class should be required. Is something wrong here?

Yes, something is wrong. Namely, due to an inadvertence of Java language, the "java.lang.String" class is defined as final. This means that no classes can be extended from it, ruining the whole beauty of OOP. In many cases Java developers need a hierarchy of classes. That hierarchy should start with a root base class, which provides basic functionality for all objects. And in many cases programming objects are very similar to strings, because they probably have a well known text representation. As a result developers are accustomed to thinking in these terms. So it would be quite natural to start the hierarchy from a string, or to extend the base root class, or just to extend a particular object from String. As we said, this is not supported in Java meaning the Java developers have to create alternative implementations of the String class and then extend all objects from that class. We implemented our CcString class entirely in Java, based on the array of characters, – an interesting exercise.

### 5.3.6 Arrays and Vectors

In section 3.6 we mentioned that our tool supports sets of beans as a special case of bean fields and interface parameters, and that the implementation of the bean sets is based on the Vector class.

Due to loose types by design, Vectors are not always convenient and their flexibility is not often used. Thus, were we to implement our tool again, we would probably use Java arrays. One of the principles of our tool is that the data types are defined in APIs and so they are exactly defined everywhere in the generated code. But Vectors also work fine and circumstances are such that we have no need to change our implementation.

### 5.3.10 JDBC Parameters

In section 3.11 we tried to design the database interface in the same pattern as our definition of the interface between a Java client and a Java server. But here we are not at the same position because JDBC limits what we can do. Firstly, the input parameters are all strings or vectors of strings. As we said we know the exact type of the parameters and can do appropriate transformations, but we cannot pass beans as parameters. Secondly, output parameters cannot be beans, although they can be vectors of beans, because the latter are returned as a cursor to a result set. Finally, before Oracle JDBC 10g, there was no support for named parameters to stored procedures. This means that the order of parameters in the stored procedure API definitions did matter.

Also we expect some cooperation from the database developer. We cannot call a stored procedure with an arbitrary signature. We require that all scalar parameters are of "varchar" or "number" types, all input database arrays are arrays of "varchars," and all returned results sets consist of "varchar" values and follow certain naming convention.

## 6. CONCLUSIONS

We offer a language independent way to define interfaces for the whole project. After that, different languages and tools can be used, each for its layer. This is different from the prevalent understanding of interfaces as tunnels to act at another layer in terms of that layer.

```
connection.prepareStatement("SELECT C1 FROM T1");

System.out.println("<TABLE><TR><TD>...");
```

In these examples Java code literally includes statements in foreign languages (SQL and HTML, correspondingly.) From our viewpoint here is no due separation of the tiers.

The proper separation of the tiers means that they can each be written in a different programming language (say, HTML, Java, and SQL,) but there should never be any mixture of the languages, or any mixture of the code belonging to different tiers in one file.

In this article we have provided an example of a GUI that is written entirely in a markup language, of an application server that is written in Java, and a back end written in SQL. Our example relies on the introduced tool, Summer MVC, facilitating development of a three tiered application, based a strict interface and data definitions.

## 7. ACKNOWLEDGEMENTS

We would like first to express our sincere appreciation to Kenneth Gaylin for the excellent job in correcting and proof reading the manuscript.

We also express our special thanks to Mr. Melvin McArthur for releasing this paper for publication.